\documentclass[]{article}

\usepackage{graphicx} % Required for inserting images
\usepackage{amsmath} % currently giving error?
\usepackage{amsfonts}
\usepackage{url}
\usepackage{amssymb}
\usepackage{hyperref}
\usepackage{color}
\usepackage{xcolor}
\usepackage[margin=1.125in]{geometry}

\begin{document}

\title{Electromagnetic Coil Optimization for Reduced Lorentz Forces}
\date{}
\author{Siena Hurwitz$^1$, Matt Landreman$^1$, Alan Kaptanoglu$^2$}
\maketitle
\vspace{-1.125cm}
\begin{center}
   $^1$Institute for Research in Electronics and Applied Physics, University of Maryland, College Park, MD 21043 USA
   
   $^2$Courant Institute of Mathematical Sciences, New York University, New York, NY 10012, USA

\end{center}

\begin{abstract}
The reduction of magnetic forces on electromagnetic coils is an important consideration in the design of high-field devices such as the stellarator or tokamak. Unfortunately, these forces may be too time-consuming to evaluate by conventional finite element modeling within an optimization loop. Although mutual forces can be computed rapidly by approximating large-bore coils as infinitely thin, this approximation does not hold for self-forces as it leads to an unphysical divergence. Recently, a novel reduced model for the self-field, self-force, and self-inductance of electromagnetic coils based on filamentary models was rigorously derived and demonstrated to be highly accurate and numerically efficient to evaluate \cite{hurwitz2023efficient}. In this paper, we present an implementation of the reduced self-force model employing automatic differentiation within the \textsc{simsopt} stellarator design software and use it in derivative-based coil optimization for a quasi-axisymmetric stellarator. We show that it is possible to significantly reduce point-wise forces throughout the coils, though this comes with trade-offs to fast particle losses and the minimum distance between coils and the plasma surface. The trade-off between magnetic forces and coil-surface distance is mediated by the minimum coil-coil distance for coils near the inboard side of the ``bean'' cross-section of the plasma. The relationship between forces and fast particle losses is mediated by the normal field error. Coil forces can be lowered to a threshold with minimal deterioration to losses. Importantly, the magnet optimization approach here can be used also for tokamaks, other fusion concepts, and applications outside of fusion.
\newline\newline
\noindent Keywords: Electromagnetic coil, Lorentz force, self-force, stellarator, nuclear fusion
\end{abstract}

\section{Introduction}
The magnetic forces exerted on electromagnetic coils frequently are important to the design of magnetic confinement fusion (MCF) reactors. Of primary concern are the engineering constraints and cost of the large structure (often of stainless steel \cite{freidberg2008plasma}) needed to support the coils against the high forces. Additionally, because future MCF devices are planned to employ powerful yet fragile high temperature superconductors (HTSs) in order to achieve higher field strengths and consequentially better confinement, another engineering consideration is the maximum allowable transverse stresses on the conductors in order to minimize fatigue \cite{freidberg2008plasma}. As an example, the modern VIPER cable architecture -- designed to withstand extreme conditions and slated for use in the SPARC tokamak \cite{hartwig2020viper} -- has been demonstrated to withstand loads of at least $300$-$400 \textrm{ kN/m}$ throughout recent experimental and numerical studies \cite{riva2023development, zhao2022structural}, a load on a similar order of magnitude or smaller than the forces exerted on coils in real-world devices \cite{hartwig2020viper}. More subtly, a third consideration is that higher magnetic forces may lead to unforeseen engineering challenges (such as fatigue on and collisions between components) due to the movements of coils under high loads. The Wendelstein 7-X stellarator in particular was plagued with a number of such challenges during its construction due to insufficient modelling capabilities and lack of resources until 2005 \cite{bykov2014specific, bykov2009structural}. 

A typical approach for determining magnetically-induced stresses and their effects upon a surrounding system is to use finite element solvers and methods such as the ANSYS software. These solvers can evaluate Maxwell's equations and other relevant models over discrete meshes and accurately model complex problems consisting of complex geometries and multiple types of material. Even so, these solvers may be difficult to set up and slow to evaluate as they typically evaluate the field throughout the volumetric mesh simultaneously as opposed to at only the points of interest. An alternative method for determining coil forces is to directly evaluate the Biot-Savart law, 
\begin{equation}
    \textbf{B}(\textbf{r})=\frac{\mu_0}{4\pi}\int \textrm{d}^3r'\frac{\textbf{J}(\textbf{r}')\times(\textbf{r}-\textbf{r}')}{|\textbf{r}-\textbf{r}'|^3},
\end{equation}
though this is also prone to slowness and numerical errors due to the integrable singularity at $\textbf{r}'=\textbf{r}$. 

Since it takes a relatively long time to calculate the forces on stellarator coils by these methods, Lorentz forces are typically considered in post-physics design rather than to inform underlying physics decisions such as through coil optimization. However, the incorporation of such considerations into the physics design itself is desirable as this would allow for a more holistic approach and presumably ease many of the aforementioned engineering challenges. It is important to note that naive attempts to speed up such calculations by making an infinitesimal thickness approximation to the coils are invalid due to  logarithmic divergences in the Biot-Savart law and self-force\cite{hurwitz2023efficient}. Such considerations however motivated recent research into simplified expressions for magnetic force. In \cite{robin2022minimization}, an expression was derived for the magnetic force on a current sheet which thus has applications to coil optimizations that rely on a winding surface approach. Additionally, in \cite{hurwitz2023efficient, landreman2023efficient}, a numerically efficient reduced model for the self-force was rigorously derived and demonstrated, with the form
\begin{equation}
    \label{eq:self_force}\frac{\textrm{d}\textbf{F}_\textrm{self}}{\textrm{d}\ell}(\phi)=I{\textbf{t}}(\phi)\times\textbf{B}_\textrm{reg}(\phi)
\end{equation}
for coils with circular or rectangular cross-sections. Here, ${\textbf{t}}$ is the tangent vector to the coil's center-line, which is an infinitesmally thin curve $\mathbf{r}_c(\phi)$ parameterized by a periodic coordinate $\phi\in[0,2\pi)$, $\textbf{B}_\textrm{reg}(\phi)$ is a \textit{regularized one-dimensional Biot-Savart law},
\begin{equation}\label{eq:reg_bs}
\textbf{B}_\textrm{reg}(\phi)=\frac{\mu_0 I}{4\pi}\oint\frac{\textrm{d}\boldsymbol{\ell}'\times(\textbf{r}_c-\textbf{r}_c')}{\left(|\textbf{r}_c-\textbf{r}_c'|^2+\delta\right)^{3/2}},
\end{equation}
and $\delta$ is a regularization factor that depends on the geometry of the coil's cross-section. (For a circular cross-section, $\delta=a^2/\sqrt{e}$, where $a$ is the minor radius and $e$ is Euler's constant. For a rectangular cross-section, $\delta$ takes on a more complex form and can be found in equations (12)-(15) of \cite{landreman2023efficient}). This model (\ref{eq:self_force},\ref{eq:reg_bs}) was shown to be highly accurate when compared with high-fidelity finite-thickness calculations but was also able to be evaluated rapidly with only $\sim 12$ quadrature points. 

A natural follow-up to this line of research -- especially in the context of stellarators -- is to use this model to optimize coils for reduced Lorentz forces. To date, one group of researchers have published brief results using this method in comparison to an optimization for the total magnetic energy of the system\cite{guinchard2024including}. In the present paper, we investigate the optimization of stellarator coils for reduced forces more thoroughly and study the trade-offs involved between forces and other key characteristics such as normalized quadratic flux and fast particle losses. In Section \ref{sec:methods}, we discuss the theory and methodology underlying this research, while in Section \ref{sec:results} we document the results and analysis. The code and data used throughout these numerical experiments is available at \url{https://doi.org/10.5281/zenodo.13913510}. 

\section{Methods\label{sec:methods}}
We performed our optimizations with the stellarator optimization suite \textsc{simsopt} \cite{landreman2021simsopt} as it has been demonstrated to allow for the achievement of precise quasisymmetry with coils \cite{landreman2022magnetic, wechsung2022precise, wiedman2023coil}. Unlike other methods such as \textsc{nescoil} in which a current potential $\Phi$ is computed on a winding surface, the filament coil representation in \textsc{simsopt} directly describes each coil shape as a Fourier expansion of a curve in real space,
\begin{equation}\label{eq:fourier}
    \textbf{r}_c(\phi)=\sum_{m=0}^\textrm{order}\textbf{a}_m\cos(m\phi)+\sum_{m=1}^\textrm{order}\textbf{b}_m\sin(m\phi).
\end{equation}
Such an approach allows a great deal of flexibility, though a downside is that the optimization space is generally non-convex. Using \textsc{simsopt}, we implemented the reduced magnetic force model  \cite{hurwitz2023efficient}. In its simplest form, the model is
\begin{equation}\label{eq:force}
    \frac{\textrm{d}\textbf{F}}{\textrm{d}\ell}(\phi)=I{\textbf{ t}}(\phi)\times\left(\textbf{B}_\textrm{reg}(\phi)+\textbf{B}_\textrm{mutual}(\phi)+\textbf{B}_\textrm{plasma}(\phi)\right),
\end{equation}
where $\textbf{B}_\textrm{reg}$ is the regularized magnetic field (\ref{eq:reg_bs}), $\textbf{B}_\textrm{mutual}$ is a filamentary approximation to the magnetic field from all external (``mutual'') coils, and $\textbf{B}_\textrm{plasma}$ is the magnetic field generated by plasma currents; we consider only vacuum fields in this paper and thus neglect $\textbf{B}_\textrm{plasma}$ for the duration. However, rather than implement (\ref{eq:reg_bs}) directly, we instead made use of the mathematically equivalent but computationally more accurate formula
\begin{equation}\label{eq:reg_mod}
    \textbf{B}_\textrm{reg}=\frac{\mu_0I}{4\pi}\int_0^{2\pi}\textrm{d}\tilde\phi\left[\frac{\tilde{\textbf{r}}_c'\times\Delta\textbf{r}}{(|\Delta \textbf{r}_c|^2+\delta)^{3/2}}-\Lambda\right]
    +\frac{\mu_0I\kappa}{8\pi}\left[-2+\ln\left(\frac{64|\textbf{r}_c'|^2}{\delta}\right)\right]\textbf{b},
\end{equation}
where $\textbf{b}(\phi)$ is the binormal unit vector to $\textbf{r}_c(\phi)$ \cite{kreyszig_1991}, $\kappa(\phi)$ is the curvature of $\textbf{r}_c(\phi)$, $\Delta$ is a notation which represents a tilde quantity subtracted from a non-tilde quantity, and
\begin{equation}
    \Lambda=\frac{\kappa(1-\cos(\Delta\phi)\textbf{b}}{2^{3/2}\left(1-\cos(\Delta\phi)+\delta/2|\textbf{r}_c'|^2\right)^{3/2}}.
\end{equation}
For performance benefits, all calculations for $\textbf{B}_\textrm{reg}$ were vectorized, and its derivatives with respect to coil shape were evaluated using the automatic differentiation package \textsc{jax} in the form of Jacobian-vector products (JVPs). This implementation is available on the main Git branch of \textsc{simsopt}. 

\begin{figure}[!htb]
\centering
\includegraphics[width=6cm]
{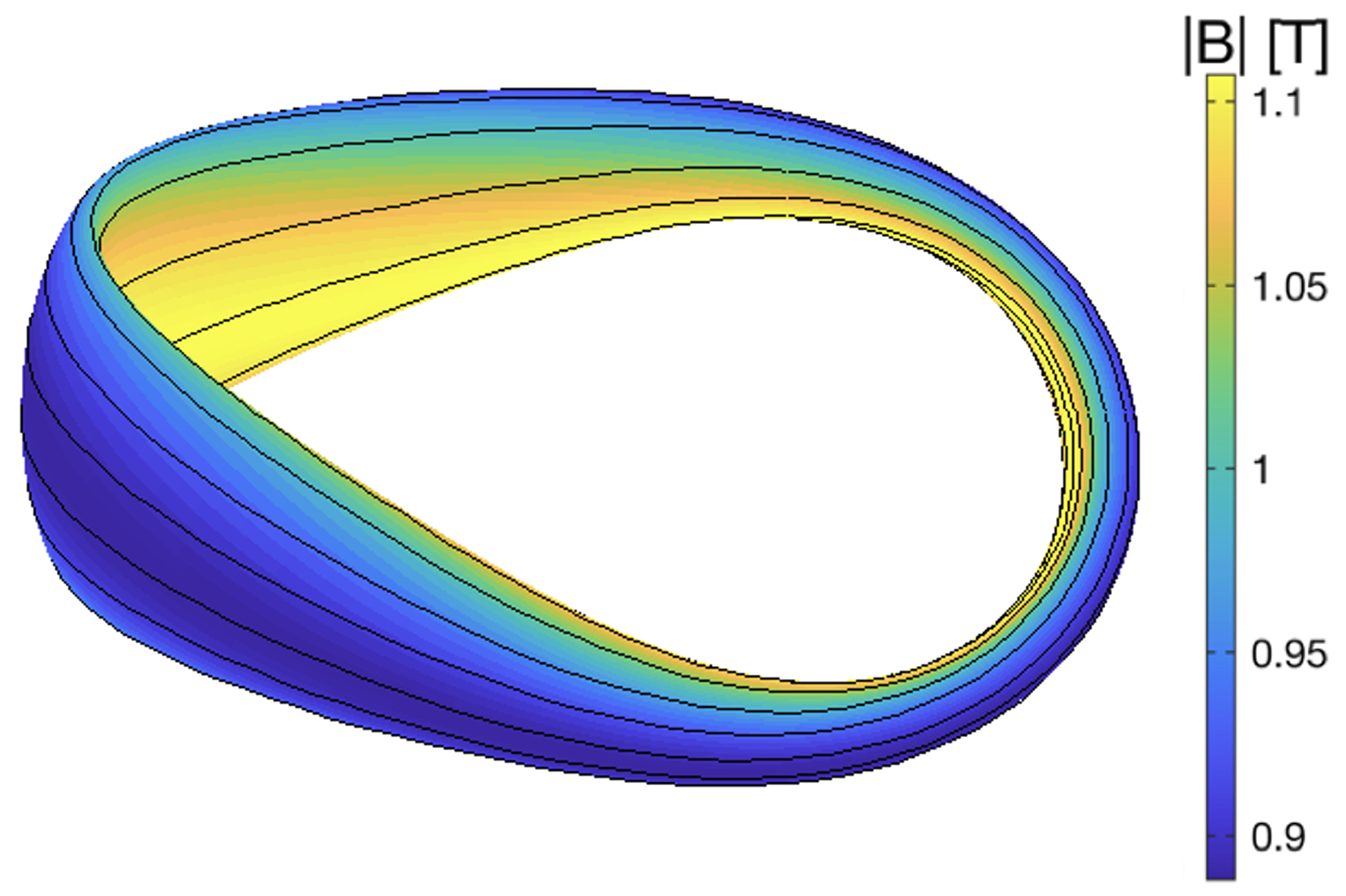}
\caption{\label{fig:QA}The precise QA configuration \cite{landreman2022magnetic}. }. 
\end{figure}

We chose to optimize coils for the highly quasi-axisymmetric vacuum equilibrium from \cite{landreman2022magnetic}, shown in Figure \ref{fig:QA}, which we hereafter refer to as the ``precise QA'' configuration. This configuration is a two field period and stellarator symmetric quasi-axisymmetric configuration for which coils were previously optimized \cite{wechsung2022precise}. 
For the context of coil optimization, we scaled the major radius to one meter. In order to maintain a reasonable balance between quasisymmetry and engineering feasibility, we decided to work with 5 coils per half field period (yielding a total of 20 coils overall) and imposed the constraints in Table \ref{tab:coil_restrictions}. In this table and throughout the remainder of the paper, we represent coil length as $\ell$, coil curvature as $\kappa$, mean-squared coil curvature as $\kappa_\textrm{MS}$, minimum distance between coils as $d_{cc}$, and minimum distance between coils and the plasma boundary as $d_{cs}$. 
\begin{table}[!ht]
\begin{center}
    
    \begin{tabular}{|l||c|c|c|c|c|} \hline 
          $R$&$\ell$&  $\kappa$&  $\kappa_{MS}$&  $d_{cc}$& $d_{cs}$\\ \hline 
          $1\textrm{ m}$&$\leq 5\textrm{ m}$&  $\leq 12\textrm{ m}^{-1}$&  $\leq 6\textrm{ m}^{-1}$&  $\geq 0.083\textrm{ m}$& $\geq 0.166\textrm{ m}$\\ \hline
 $7.75\textrm{ m}$& $\leq 38.9\textrm{ m}$& $\leq 1.55\textrm{ m}^{-1}$& $\leq 0.77\textrm{ m}^{-1}$& $\geq 0.646\textrm{ m}$&$\geq 1.29\textrm{ m}$\\\hline
    \end{tabular}
    \caption{This table lists the constraints imposed upon coil sets generated for this paper. The first row corresponds to a plasma scaled to a one meter major radius (the scale used for optimization), while the second row gives the equivalent values after scaling to the minor radius of the ARIES-CS reactor. \label{tab:coil_restrictions}} 
\end{center}
\end{table}

We designed our objective function $f$ to penalize deviations from the target equilibrium as measured by the normal field error; averaged point-wise Lorentz forces; coil arc-length variation to avoid poor conditioning \cite{bindel2023understanding, wechsung2022precise}; and various engineering and regularization terms such as coil length, minimum coil-coil distance, minimum coil-surface distance, coil curvature, and mean-squared curvature. We write this objective function as
\begin{equation}\label{eq:objective}
    f=f_\textrm{qf}+f_\textrm{F}+f_\sigma+f_\ell+f_\textrm{cc}+f_\textrm{cs}+f_\kappa+f_{\kappa_\textrm{MS}},
\end{equation}
where
\begin{subequations}
    \begin{align}
        f_\textrm{qf}=\frac{w_\textrm{qf}}{2}\int_S\textrm{d}S(\textbf{B}\cdot\textbf{n})^2\\
        f_\textrm{F}=\frac{w_\textrm{F}}{2}\sum_\textrm{coils}\oint\textrm{d}\ell\max(|\textrm{d}\textbf{F}/\textrm{d}\ell|-|\textrm{d}\textbf{F}/\textrm{d}\ell|_0,0)^2\\
        f_\sigma=w_\sigma\sum_\textrm{coils}\textrm{Var}(\textrm{d}\ell)\\
        f_\ell=w_\ell\sum_\textrm{coils}(\ell-\ell_0)^2\\
        f_\textrm{cc}=\sum_\textrm{coil pairs}\max(d_\textrm{cc,0}-d_\textrm{cc},0)^2\\
f_\textrm{cs}=w_\textrm{cs}\sum_\textrm{coils}\max(d_\textrm{cs,0}-d_\textrm{cs},0)^2\\
        f_\kappa=\frac{w_\kappa}{2}\sum_\textrm{coils}\oint\textrm{d}\ell\max(\kappa-\kappa_0,0)^2\\
f_{\kappa_\textrm{MS}}=w_{\kappa_\textrm{MS}}\sum_\textrm{coils}\max\left(\left(\frac{1}{\ell}\oint \textrm{d}\ell \kappa^2\right) - \kappa_\textrm{MS,0},0\right)^2,
    \end{align}
\end{subequations}
$w$'s represent the various weights, and all quantities with a subscript of zero indicate the target values in Table \ref{tab:coil_restrictions}. We also arbitrarily chose the regularization factor $\delta$ of the regularized Biot-Savart law (\ref{eq:reg_bs}) to correspond to that of a circular cross-section with a minor radius of $0.05$ meters. 

\section{Results\label{sec:results}}
\subsection{Optimizations\label{sec:optimizations}}
\begin{figure}[!htb]
\centering
\includegraphics[width=8cm]
{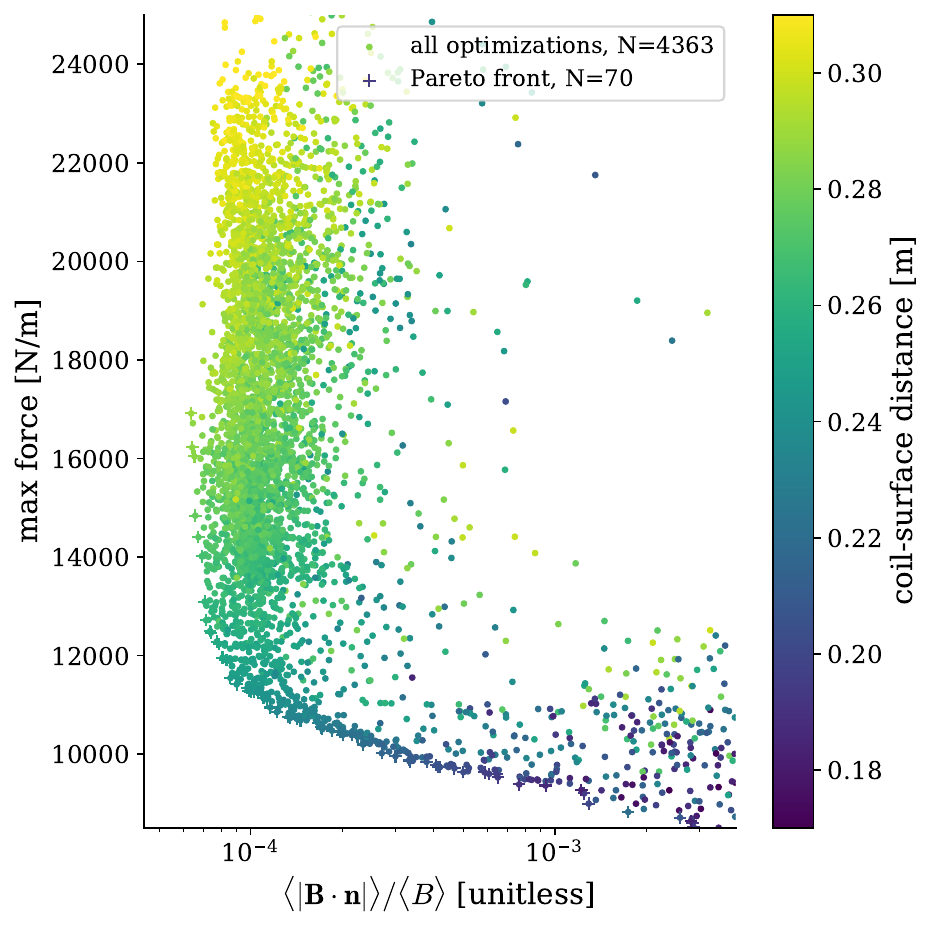}
\caption{\label{fig:cold_start}This plot shows the cold start coil optimizations in the space of maximum point-wise force (\ref{eq:max_f}) versus normal field error (\ref{eq:norm_flux}). In the legend, $N$ represents the number of optima in each category.}
\end{figure}

We began with a set of ``cold start'' optimizations in which the coils were initialized to be a set of equidistant circles lying on the surface of a torus. Formally, we describe the optimizations as seeking the solution $\textbf{x}_0$ to the problem
\begin{equation}
    \min_{\textbf{x}_0}f(\textbf{x}_0;\boldsymbol{\alpha}_0),
\end{equation}
where $\boldsymbol{\alpha}_0$ is the set of hyperparameters (the cost function weights and initial conditions for the cold start) and $\textbf{x}_0$ is the vector of degrees of freedom for the problem (the Fourier coefficients (\ref{eq:fourier}) and the coil currents). For each optimization, a unique objective function was determined through a combination of hyperparameters randomly chosen from a predetermined range that tended to yield results in alignment with Table \ref{tab:coil_restrictions}. We chose to use such an approach of sampling a large space in order to approximate the global minimum as this type of optimization problem generally suffers from multiple local minima. Once the objective function was initialized, coil optimization was performed using the limited-memory BFGS algorithm for its speed of convergence. Upon completion of the cold start optimizations, a set of filters were applied in order to exclude coil sets which did not meet the criteria in Table \ref{tab:coil_restrictions} or were linked and/or detached from the plasma surface. The results of the filtered cold start optimizations are shown in Figure \ref{fig:cold_start}, where each optimization is plotted in the space of maximum point-wise force across all coils
\begin{equation}\label{eq:max_f}
    \max_\textrm{coils}\left(\max_\phi \bigg|\frac{\textrm{d}\textbf{F}}{\textrm{d}\ell}\bigg|\right)
\end{equation}
against normalized normal field error over the last closed flux surface (LCFS)
\begin{equation}
    \frac{\langle|\textbf{B}\cdot\textbf{n}|\rangle}{\langle B\rangle},\label{eq:norm_flux}
\end{equation}
where the angle brackets $\langle\cdots\rangle$ denote a flux surface average and the normal field error is a measure of deviation of the optimized solution from the target surface. 
Here, the color indicates the minimum coil-plasma surface distance and demonstrates its strong correlation with the maximum force, a relationship discussed in further detail later. We also plotted the Pareto front for these optimizations in order to visualize the trade-off involved in these optimizations, where each Pareto optimum describes a solution for which no other optimum is better in at least one regard \cite{bindel2023understanding}. For this reason, the Pareto front is generally a helpful tool for understanding trade-offs.
While there is not a single ideal optimum in a problem such as this with multiple objectives, it can still be said that points not on the Pareto front are otherwise inferior. 

\begin{figure}[!htb]
\centering
\includegraphics[width=7cm]
{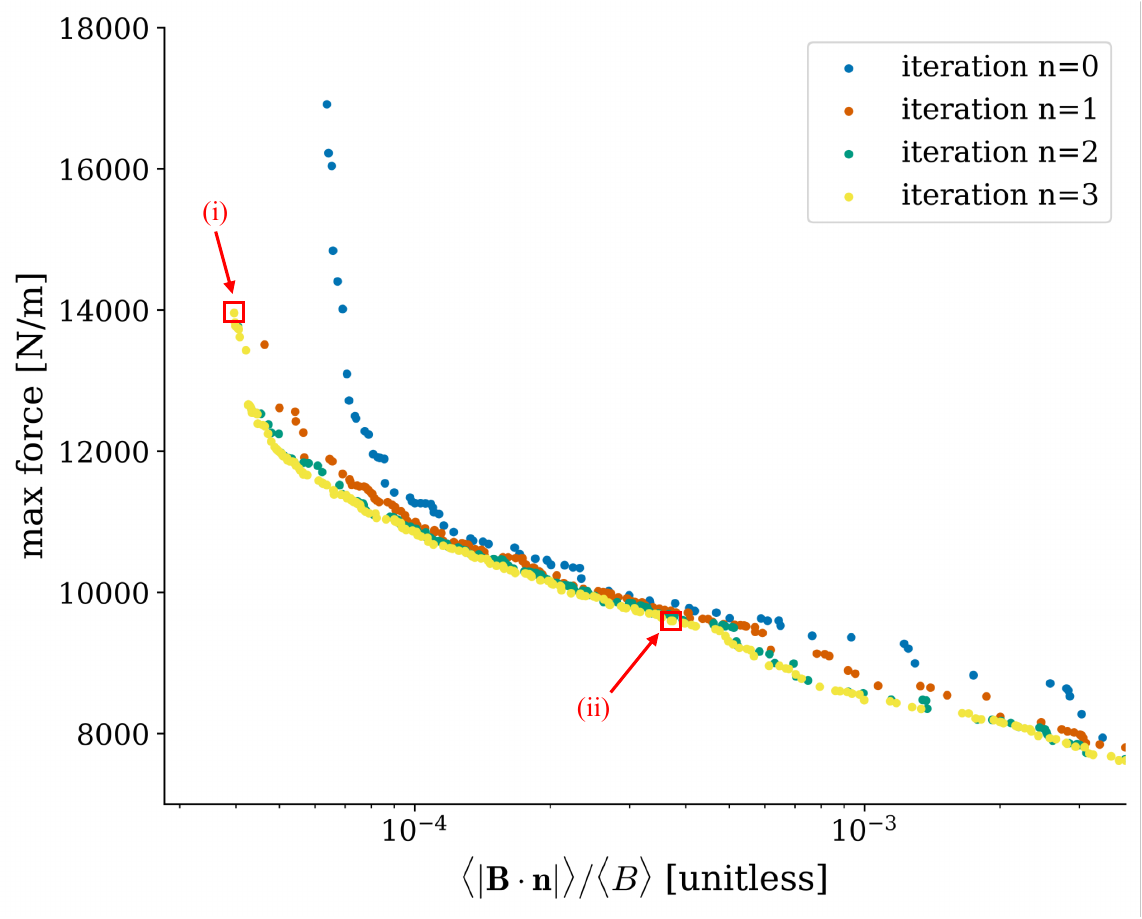}
\caption{\label{fig:continuation}This plot shows the Pareto fronts for the cold start ($n=0$) and the hot starts ($n=1, 2, 3$). For future illustrative purposes, we also select two optima we denote as (i) and (ii).}
\end{figure}

\begin{figure}[!htb]
\centering
\includegraphics[width=4cm]
{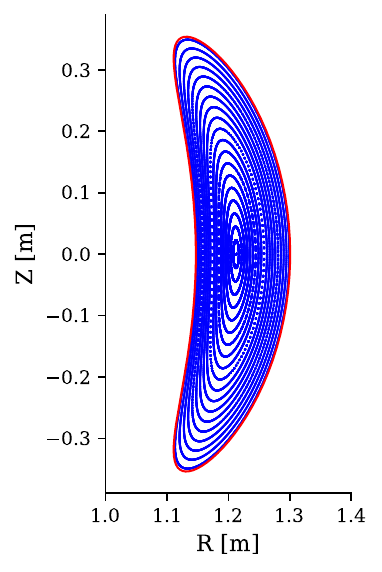}
\caption{\label{fig:poincare}The target LCFS (red) and Poincaré plot (blue) at the toroidal slice $\zeta=0$ of the field generated by coils from optimum (ii) of Figure \ref{fig:continuation}. }

\end{figure}

\begin{figure}[!htb]
\centering
\includegraphics[width=7cm]
{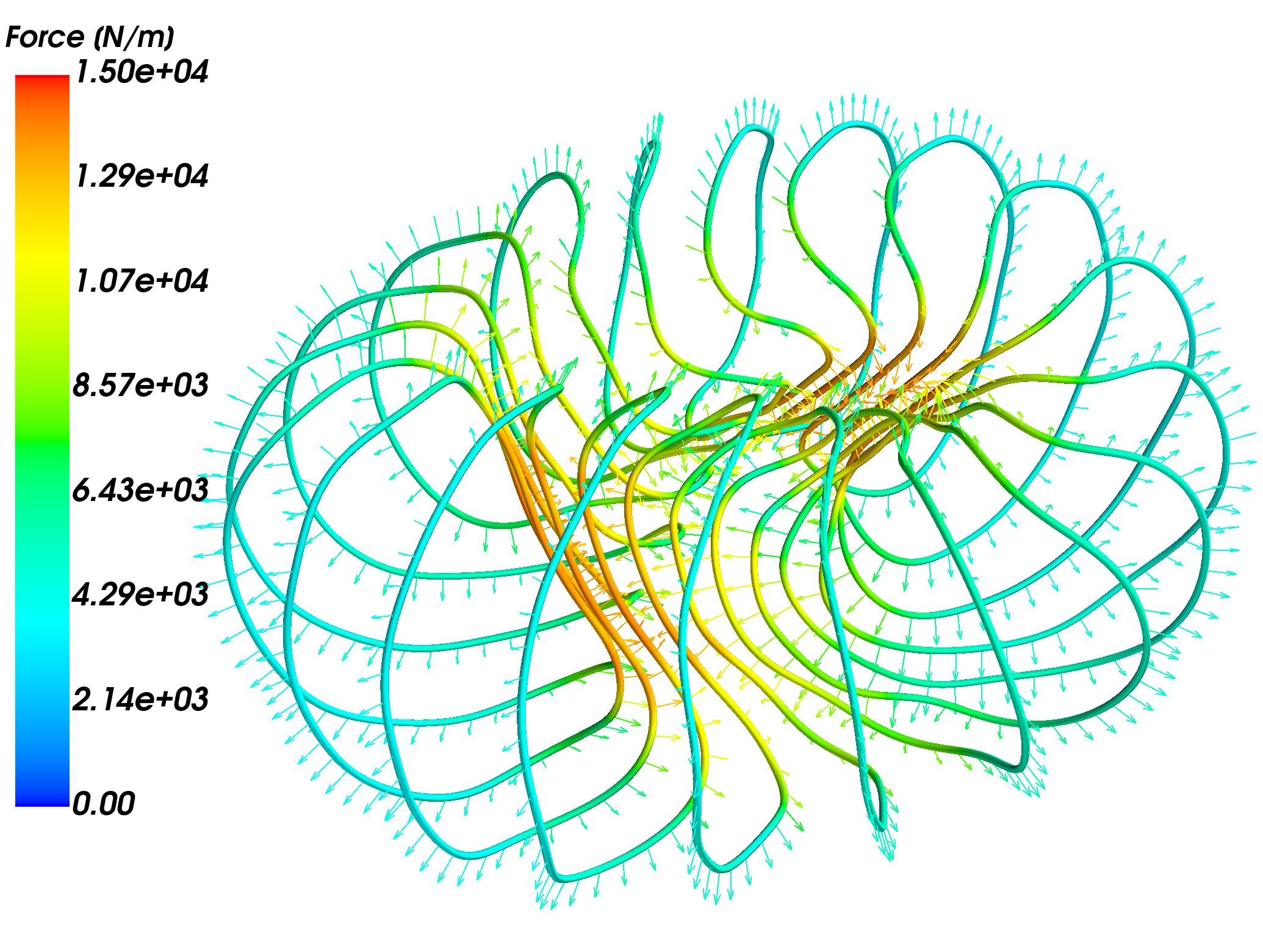}\includegraphics[width=7
cm]
{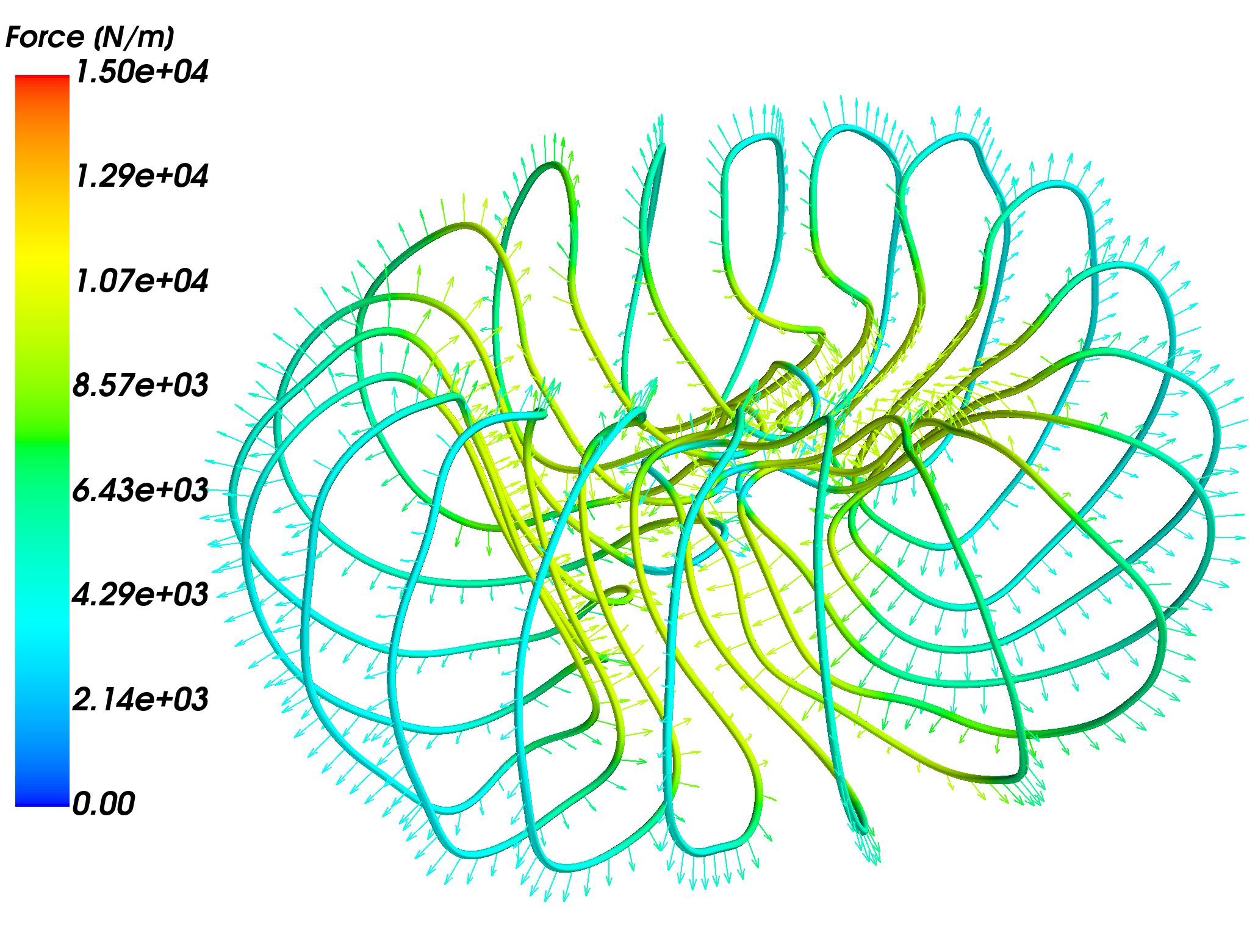}
\caption{\label{fig:coils}A visual comparison of coils noted in Figure \ref{fig:continuation}, where (i) (left) corresponds to the optimum with higher force and lower normal field error, and (ii) (right) corresponds to the optimum with lower force and higher normal field error. Note that the magnitude of the force is only indicated by the color and not by the length of the arrows.}

\end{figure}

After the initial set of cold start optimizations was completed, we performed a further series of optimizations using a continuation method in order to explore and expand the original Pareto front \cite{bindel2023understanding}. Generally, continuation methods describe methodologies in which a new optimization problem is solved by perturbing a previous one and using the previous problem's solution as an initial guess. So long as the deformation is small enough, we can expect to generate a new series of solutions similar to yet different from the earlier solutions. Here, we began each $n$th iteration by taking our initial guesses for the coil shapes to be solutions along the $(n-1)$th Pareto front. For this reason, we also term the continuations in this paper ``hot starts'' in contrast to the initial cold starts. Formally, our continuation method then represents the solutions $\textbf{x}_n$ to the problem
\begin{equation}
    \min_{\textbf{x}_n}f(\textbf{x}_n;\boldsymbol{\alpha}_n),
\end{equation}
where
\begin{equation}
    \boldsymbol{\alpha}_n=\boldsymbol{\alpha}_{n-1}+\Delta\boldsymbol{\alpha}
\end{equation}
is some small perturbation to the hyperparameters of the $n-1$ solution; generally, we choose each element of the perturbation to be less than 5\% of the original value. We plot the Pareto results of our cold and hot starts in Figure \ref{fig:continuation} and select for visualization purposes two optima we denote as (i) and (ii). 
In Figure  \ref{fig:poincare}  we show a Poincaré plot of (ii) to demonstrate that the target boundary is reproduced well with coils. Poincaré plots of other optima show similarly nested surfaces.
In Figure \ref{fig:coils}, we show plots of the coils for (i) and (ii), where the color-bar represents the point-wise Lorentz force magnitude. 

\begin{figure}[!htb]
\centering
\includegraphics[width=\textwidth]
{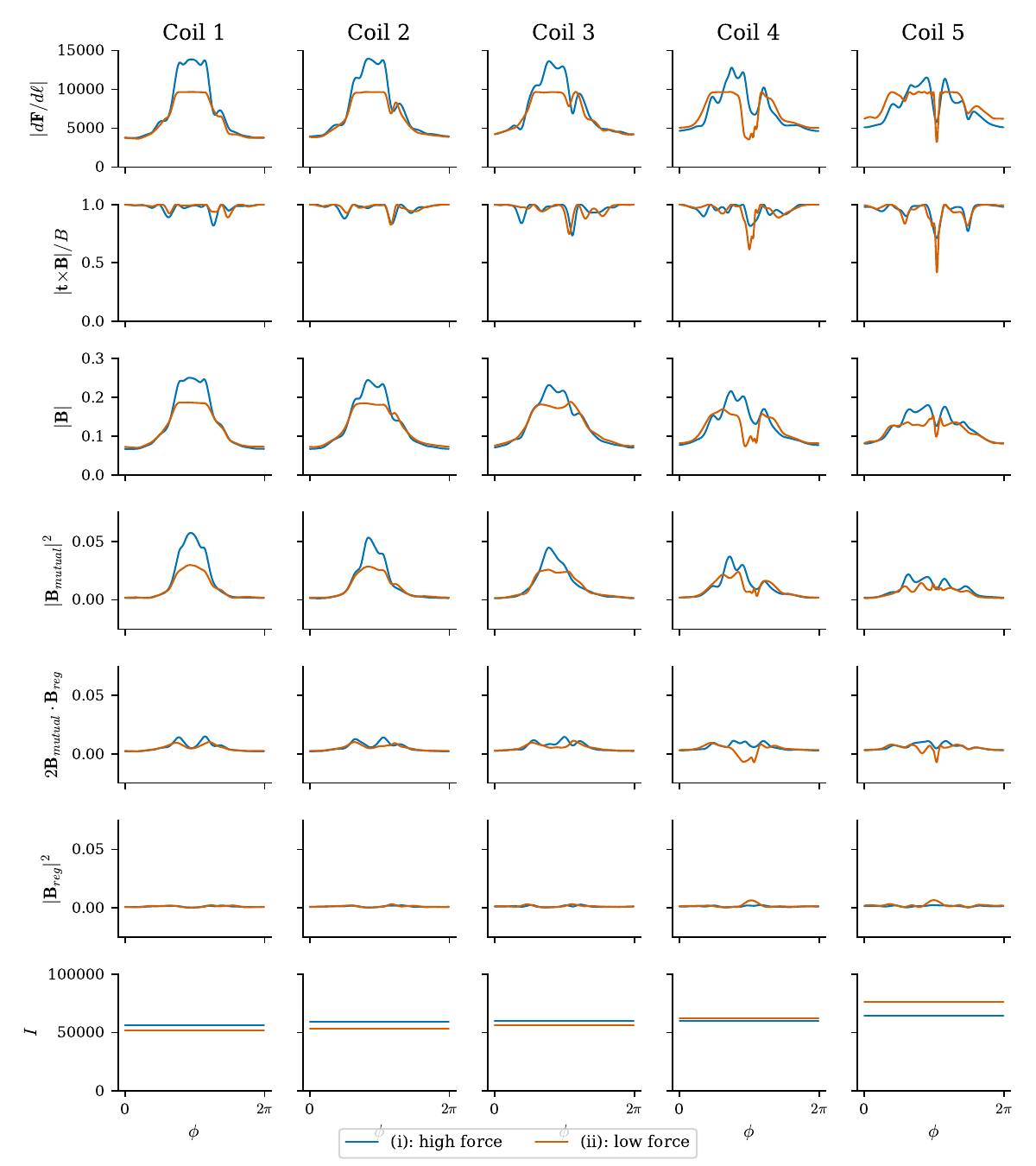}
\caption{\label{fig:force_breakdown}Various components of the point-wise force (\ref{eq:force}) plotted against curve parameter for all coils in the sets (i) and (ii). The units for all axes are standard SI units.}

\end{figure}

It is insightful to analyze how exactly the optimizer reduces coil force. From (\ref{eq:force}), we note that the product $\textbf{t}\times\textbf{B}$ can be decomposed into a scalar $B=\sqrt{B_\textrm{reg}^2+B_\textrm{mutual}^2+2\textbf{B}_\textrm{reg}\cdot\textbf{B}_\textrm{mutual}}$ 
multiplied by a cross product $\textbf{t}\times(\textbf{B}/B)$ of unit vectors. In this sense, the force at any one point can be reduced through a combination of five separate actions: (1) reducing the current in a coil, (2) aligning the tangent vector $\textbf{t}$ of the coil with the magnetic field, (3) reducing the magnitude of $\textbf{B}_\textrm{reg}$, (4) reducing the magnitude of $\textbf{B}_\textrm{ext}$, or (5) reducing the dot product of $\textbf{B}_\textrm{reg}$ and $\textbf{B}_\textrm{ext}$. In order to understand which actions are predominant, we show in Figure \ref{fig:force_breakdown} a deconstruction of the various components of (\ref{eq:force}) for each coil in the sets (i) and (ii). 
The first takeaway is that the optimizer focuses less on reducing $(\textbf{t}\times\textbf{B})/B$ than it does on reducing the strength of the field $B$ itself. The greatest reductions in the strength of the field occur on the inboard side of the coils, corresponding to curve parameter $\phi\approx\pi$. As seen in Figures \ref{fig:force_breakdown} and \ref{fig:coils}, it is here that the forces—specifically \textit{mutual} forces—are highest. This also explains why we have seen previously (figure \ref{fig:cold_start}) a strong and positive correlation between forces and the minimum coil-surface distance: near the inboard mid-plane the coils are most densely clustered, so they have the least ability to reduce their mutual forces except by moving away from one another and towards the LCFS. Such a behavior can be seen in Figure \ref{fig:coils}. For this reason, it may be beneficial to include a penalty term on small magnetic field scale length $L_{\nabla B}$ during stage-one optimizations, as recent research showed a strong positive correlation between $L_{\nabla B}$ and minimum coil-surface distance \cite{kappel2024magnetic}. Paradoxically, the minimum coil-coil distance does not display a strong correlation either with maximum force ($R^2$=0.09) or the minimum coil-surface distance ($R^2=0.17$). The reason for this is that the region where the force is the highest (on the inboard side at toroidal angles $\zeta \approx 0$ and $\pi$ where the plasma cross-section is a ``bean'') is not the same region where the coils are closest together (at toroidal angles $\zeta \approx \pi/2$ and $3\pi/2$ where the plasma cross-section is a ``triangle''). Therefore, the minimum coil-coil distance near $\zeta \approx 0$ (specifically amongst coils 1 and 2 from Figure \ref{fig:force_breakdown}) as opposed to the global minimum coil-coil distance can thus be understood to mediate the positive correlation between coil-surface distance and force, as this quantity shows a much stronger correlation with minimum coil-surface distance ($R^2=0.44$).  

Another key takeaway is that, while the mutual field dominates the total force on the coils, the self-field is non-negligible. Although $\textbf{B}_\textrm{reg}$ itself is small enough that the optimizer sometimes chooses to \textit{increase} its magnitude, reducing the vector product $\textbf{B}_\textrm{mutual}\cdot\textbf{B}_\textrm{reg}$ can have meaningful effects, such as seen with coil 4 in Figure \ref{fig:force_breakdown}. Finally, we note that the freedom to vary the individual coil currents $I$ plays a role as the optimizer chooses to redistribute the currents away from the coils experiencing the highest forces and towards the coils experiencing the lowest forces.

\subsection{Analysis\label{sec:field_analyses}}
\begin{figure}[!htb]
\centering
\includegraphics[width=8cm]
{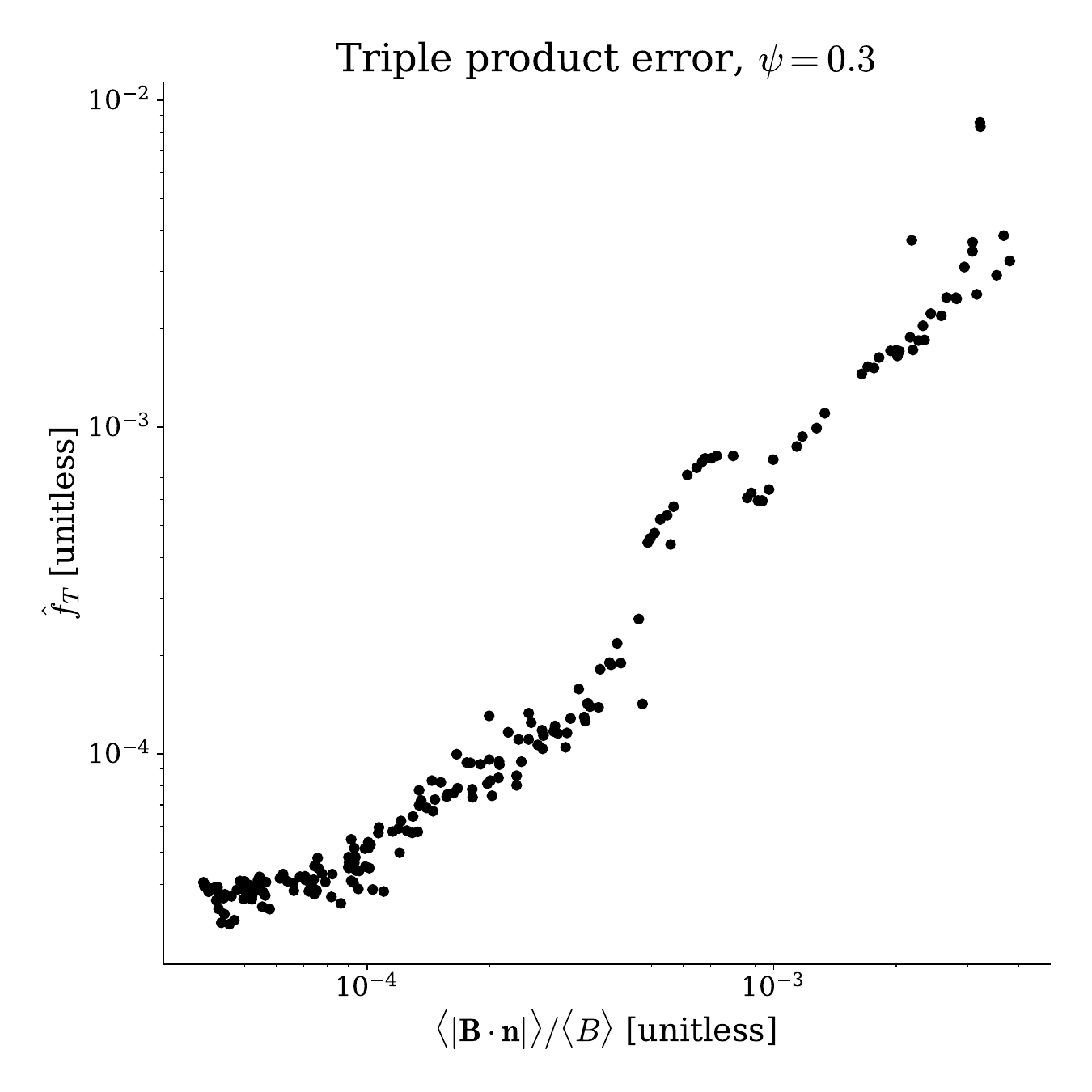}
\caption{\label{fig:fT_vs_BdotN}A comparison of the averaged quasisymmetry triple product error (\ref{eq:triple_prod_mod}) with normal field error for the final Pareto front shown in Figure \ref{fig:continuation}.}
\end{figure}

\begin{figure}[!htb]
\centering
\includegraphics[width=6.5cm]
{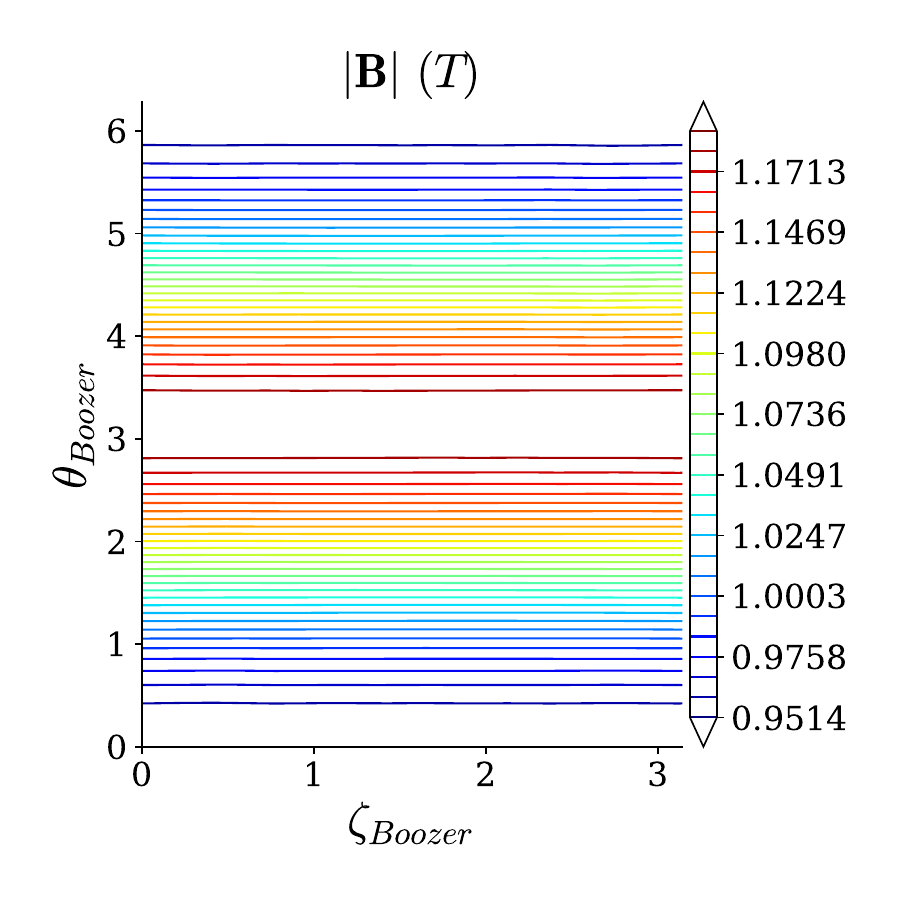}\includegraphics[width=6.5cm]
{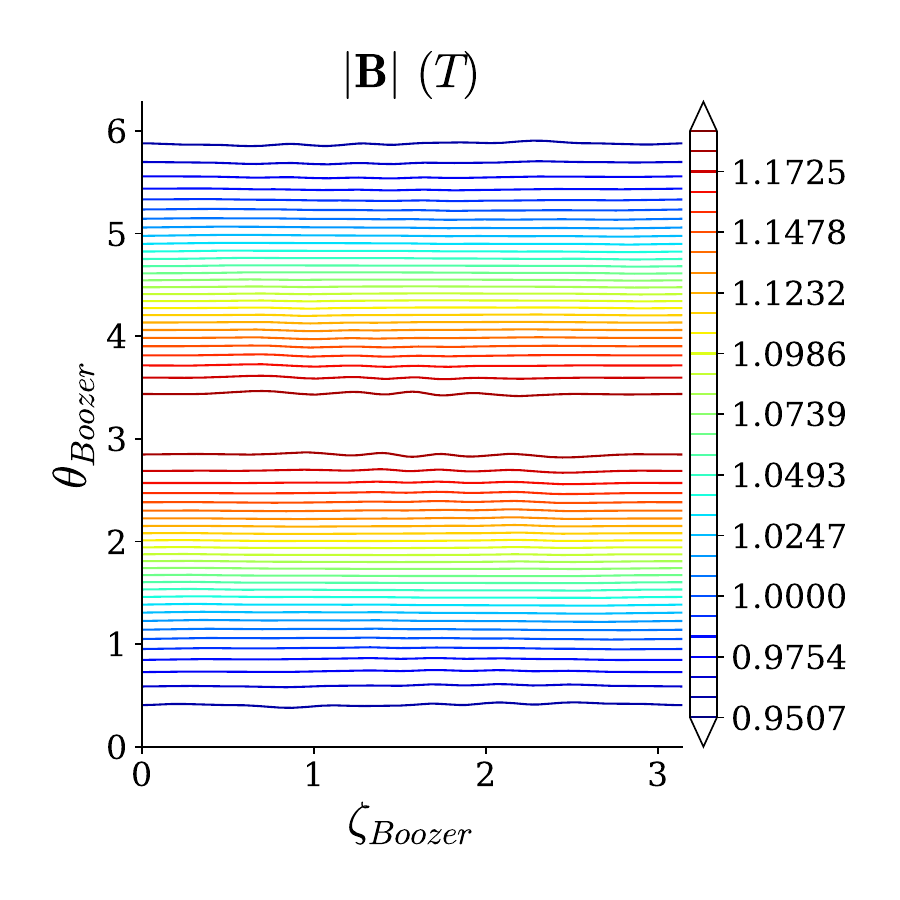}
\caption{\label{fig:|B|}Plots of $|\textbf{B}|$ in Boozer coordinates for coils  from Figure \ref{fig:continuation}, where (i) (left) corresponds to the equilibrium with higher Lorentz force and lower normal field error, and (ii) (right) corresponds to the optimum with lower force and higher normal field error. The $B$ contours are evaluated at the free-boundary flux surface with equal volume to the boundary of the target configuration.}
\end{figure}

The target equilibrium is highly quasisymmetric, so we wished to demonstrate that the coils similarly produce a quasisymmetric field. There are three common metrics for quasisymmetry error: (1) the strength of symmetry-breaking modes in Boozer coordinates, (2) a two-term error, and (3) the triple product error \cite{rodriguez2022measures,dudt2023desc}. Here, we use the triple product error
\begin{equation}
    f_T=\nabla \psi \times \nabla B \cdot \nabla(\textbf{B} \cdot \nabla B)\label{eq:triple_prod}
\end{equation}
as it is a local representation of quasisymmetry error and may best predict particle transport of the three options \cite{dudt2023desc}. More precisely, as the triple product error varies over a flux surface, we use a normalized and flux surface-averaged error
\begin{equation}
    \hat{f}_T=\frac{\langle R\rangle^2\langle|f_T|\rangle}{\langle B\rangle^4},\label{eq:triple_prod_mod}
\end{equation}
where $\langle R\rangle$ is the averaged major radius. In Figure \ref{fig:fT_vs_BdotN}, we plot the triple product error $\hat{f}_T$ against normal field error for the final iteration of the continuation method and, for consistency with the upcoming fast particle tracing calculations, evaluate the triple product error at the flux surface with normalized toroidal flux $\psi=0.3$. 
As expected, the normal field error displays a strong positive relationship with quasisymmetry error. We also show plots of $|\textbf{B}|$ contours in Boozer coordinates in Figure \ref{fig:|B|} as these contours form straight lines under perfect quasisymmetry. 
These plots allow us to verify that the coil sets achieve excellent quasisymmetry as the field contours are nearly straight in both cases. Additionally, we see that some coil ripple is visible for case (ii), which is indicative of the smaller coil-to-plasma distance and higher normal field error.

\begin{figure}[!htb]
\centering
\includegraphics[width=8cm]
{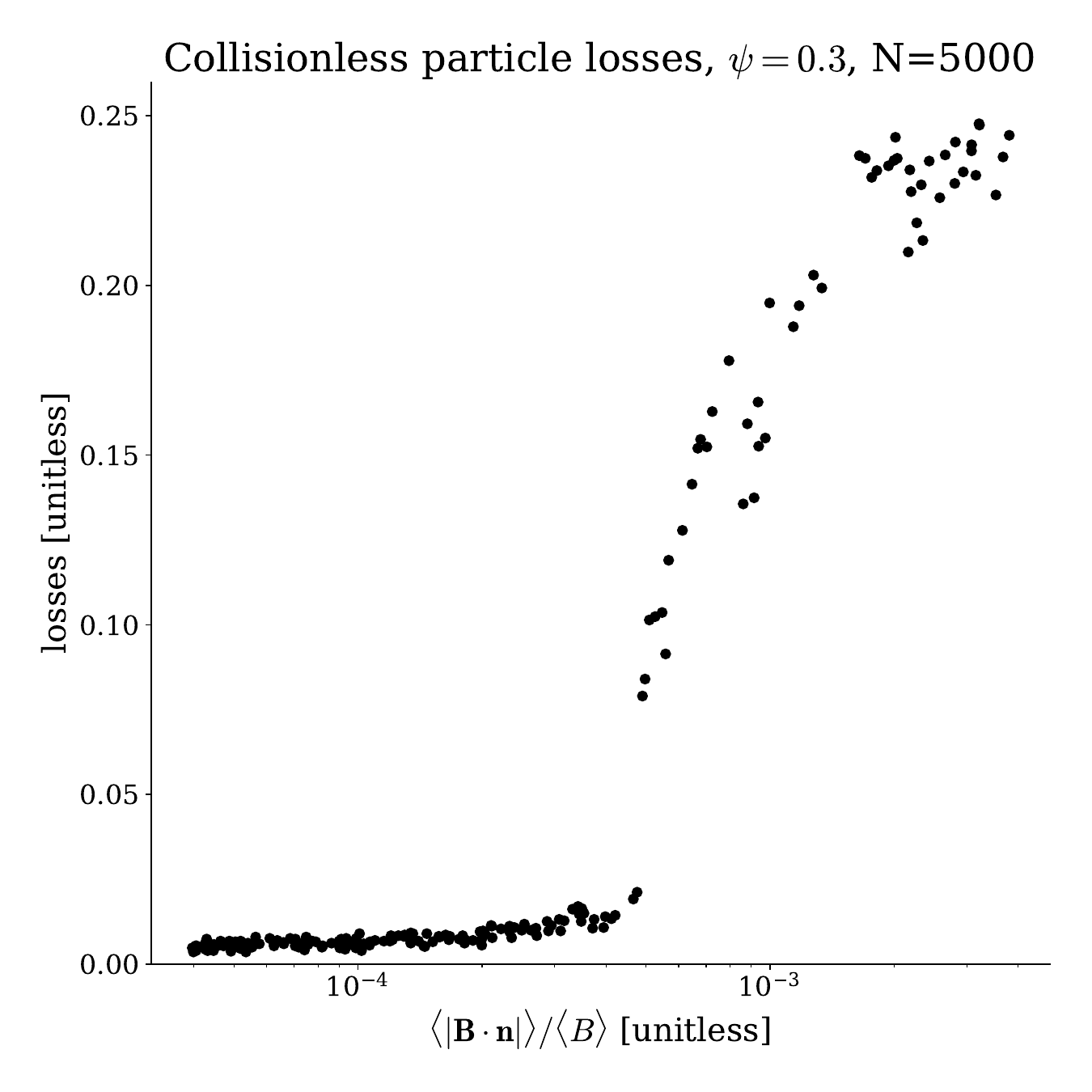}
\caption{\label{fig:losses_vs_BdotN}A comparison of collisionless fast particle losses after $0.2$ seconds for 5000 $\alpha$-particles launched from the $\psi=0.3$ normalized toroidal flux surface with normal field error for the final Pareto front shown in Figure \ref{fig:continuation}.}
\end{figure}

\begin{figure}[!htb]
\centering
\includegraphics[height=7cm]
{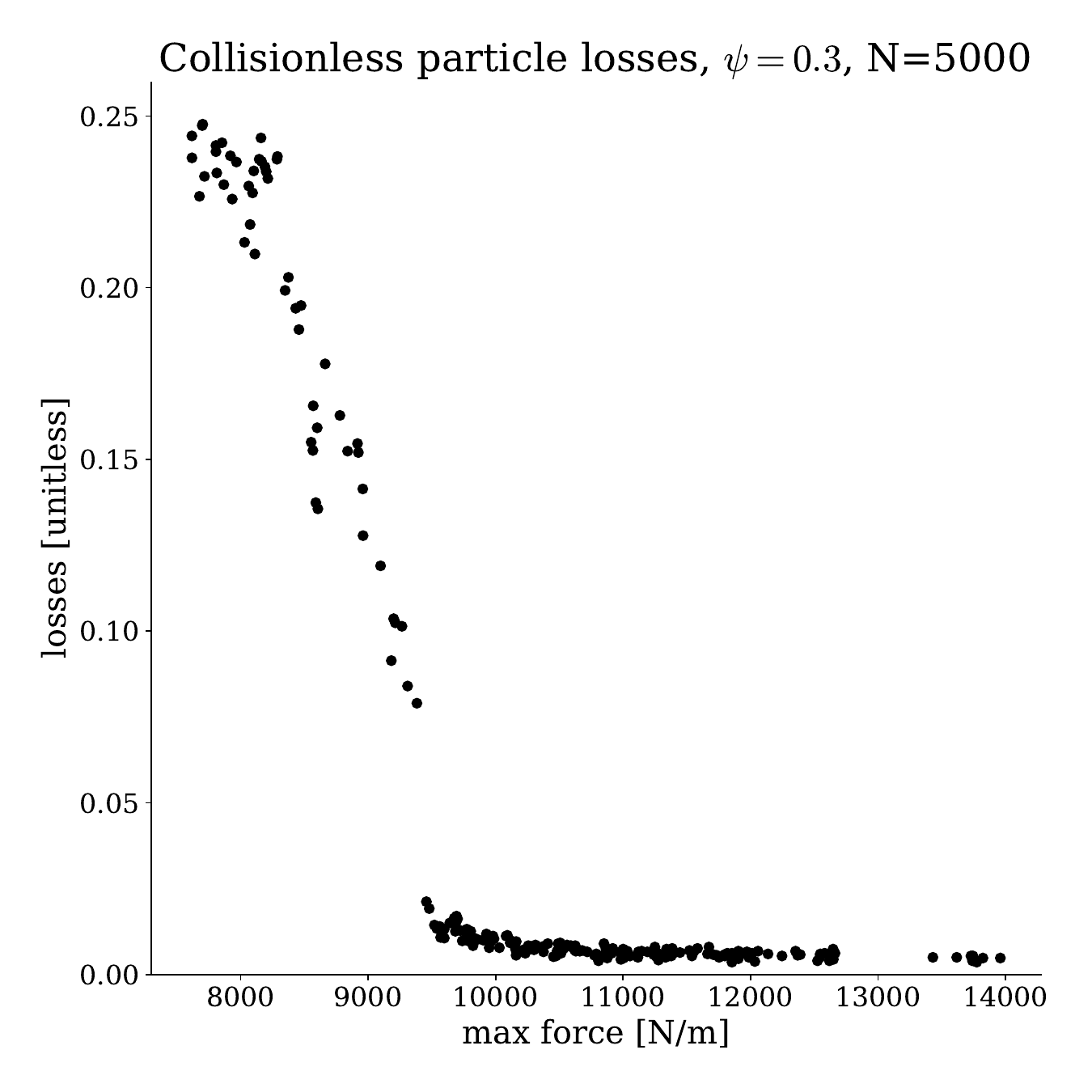}
\includegraphics[height=7cm]
{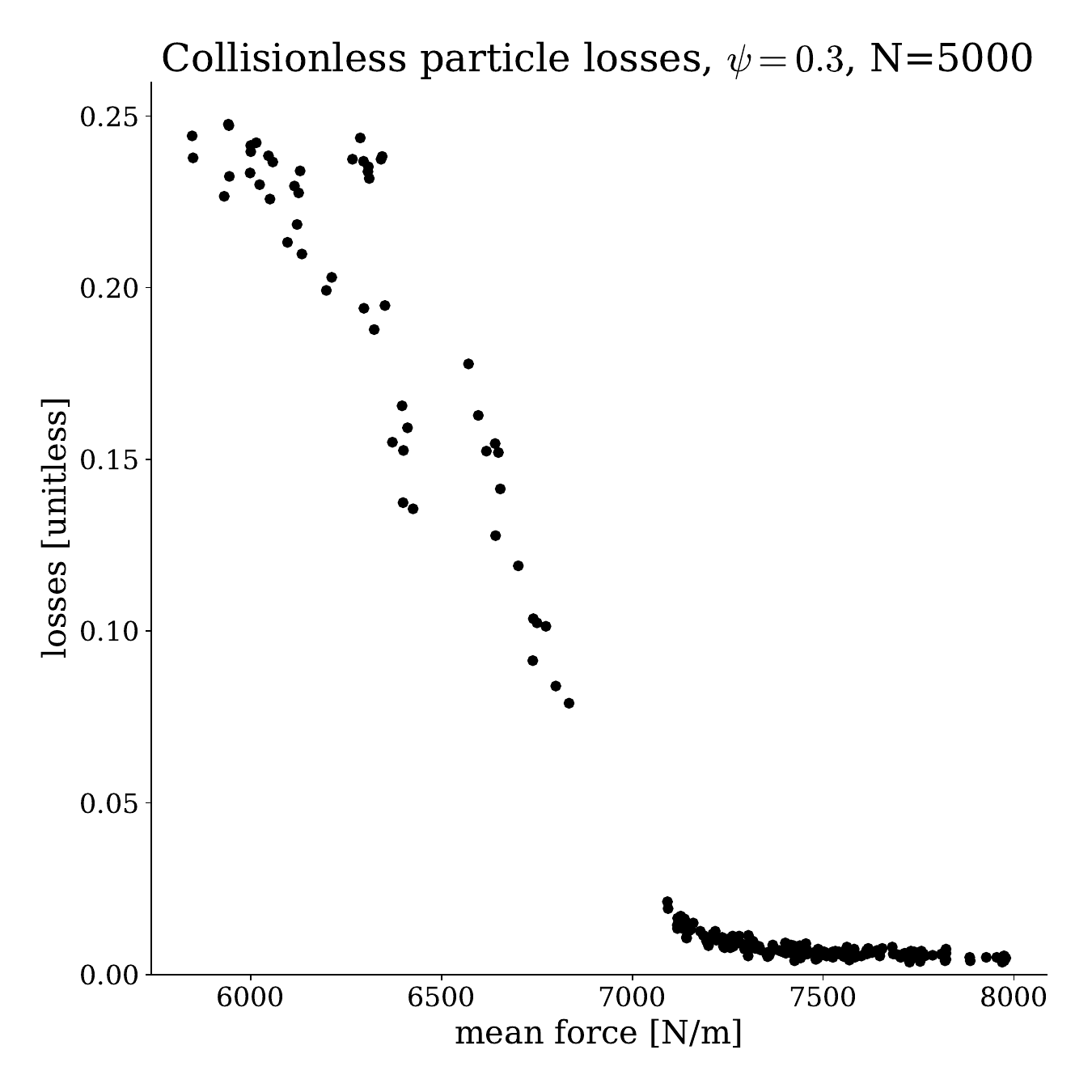}
\caption{\label{fig:losses_vs_force}A comparison of collisionless fast particle losses after $0.2$ seconds for 5000 $\alpha$-particles launched from the $\psi=0.3$ normalized toroidal flux surface with (a) maximum point-wise force (\ref{eq:max_f}) and (b) mean point-wise force (\ref{eq:f_mean}) for the final Pareto front shown in Figure \ref{fig:continuation}. Note that the fast particle losses were calculated on the reactor-scale plasma while the forces were calculated on the original configuration.}
\end{figure}

We next performed particle tracing using the guiding-center code SIMPLE \cite{albert2020accelerated} in order to investigate fast particle losses. Similarly to \cite{wechsung2022precise} and \cite{wiedman2023coil}, we scaled the minor radius and volume-averaged magnetic field strength of our configuration to the reactor-scale ARIES-CS \cite{najmabadi2008aries}, from which we launched $5000$ $3.5\textrm{ MeV}$ alpha particles at the $\psi=0.3$ normalized toroidal flux surface and traced their trajectories for $0.2$ seconds. As seen in Figure \ref{fig:losses_vs_BdotN}, the loss fraction is closely and positively correlated with the normal field error and is as small as $0.36\%$ for some optima. 
Additionally and as seen in Figure  \ref{fig:losses_vs_force}, the losses show a negative correlation with the maximum (\ref{eq:max_f}) and mean forces across the coils, where the mean force is defined as
\begin{equation}\label{eq:f_mean}
    \frac{1}{N_\textrm{coils}}\sum_\textrm{coils}\sqrt{\frac{1}{\ell}\oint\textrm{d}\ell\bigg|\frac{\textrm{d}\textbf{F}}{\textrm{d}\ell}\bigg|^2}.
\end{equation}
Notably, the losses remain small at higher forces, and only begin to significantly deteriorate below a certain threshold (approximately $9500\textrm{ N/m}$ for the maximum force and $7000\textrm{ N/m}$ for the mean force). This has the important implication that forces in these configurations can be significantly reduced—to a point—with only minimal trade-offs in fast particle confinement. 

\section{Conclusion}
Building upon prior research \cite{hurwitz2023efficient, landreman2023efficient} that derived a highly accurate and numerically efficient reduced model for electromagnetic coil self-forces, we implemented this model in the stellarator optimization suite \textsc{simsopt}. By differentiating through the reduced model using automatic differentiation, gradient-based optimization was used to optimize coils with reduced Lorentz forces for a precisely quasi-axisymmetric configuration \cite{landreman2022magnetic}. These coils accurately reproduce the target flux surface and achieve good levels of quasisymmetry and fast particle loss; the best equilibrium achieves losses of only $0.36\%$ in $0.2$ seconds when alpha particles were launched from the $\psi=0.3$ surface. We also found that it is possible to significantly reduce the peak Lorentz forces, albeit with a number of important trade-offs. First, coil forces are strongly correlated with the minimum distance between the coils and plasma since the optimizer attempts to move coils on the inboard side closer to the plasma in order to increase the coil-coil distances there and hence reduce large mutual forces. This behavior presumably explains the strong negative correlation between coil forces and normal field error: decreasing coil forces leads to a lower coil-surface distance, which therefore increases the normal field error due to coil ripple. This in turn causes the trade-off seen between forces and fast particle losses. Notably, though a reduction in forces does lead to worse particle confinement, the losses remain minimal until the maximum forces drop below a certain threshold, indicating that there is flexibility up to a point to reduce coil forces with minimal trade-offs.

In this study we minimized the magnitude of the point-wise force density $|\textrm{d}\textbf{F}/\textrm{d}\ell|$, though in future work it will be important to determine the most meaningful quantities to optimize for and how to incorporate models for the mechanical properties of coils or the support structure. For example, it may be helpful to reduce lateral forces to simplify support structures, reduce shear in the forces, minimize the total force $\textbf{F}$ on each coil, minimize torque on the coils, or reduce the mechanical strain. Any of these objectives could be computed efficiently using the same reduced model for Lorentz forces (\ref{eq:force}, \ref{eq:reg_mod}). Additionally, it would be helpful to examine the similarities and differences between the results for the quasi-axisymmetric configuration here with results for other stellarator configurations such as quasi-helically symmetric or quasi-isodynamic plasmas. Moreover, a similar optimization method could be applied to magnets in other fusion concepts such as tokamaks, or to other technologies that rely upon high-field magnets such as magnetic resonance imaging. 

\section*{Acknowledgments}
We thank David Bindel, Stefan Buller, Rory Conlin, Nicolò Foppiani, Paul Huslage, Emily Ingram, Jonathan Kappel, Thomas Klinger, and Maximilian Ruth for useful discussions. This work was supported by a grant from the Simons Foundation (No. 560651, T. A.) and by the U.S. Department of Energy under Contract DE-FG02-93ER54197. This research used resources of the National Energy Research Scientific Computing Center (NERSC), a Department of Energy Office of Science User Facility using NERSC award FES-ERCAP-mp217 for 2024. Matt Landreman is a consultant for Type One Energy Group.

\bibliographystyle{plain}
\bibliography{main.bib}

\end{document}